\documentclass{aa}
\usepackage{graphicx}
\usepackage{times}

\begin{document}

\title{PX Andromedae: Superhumps and variable eclipse depth\thanks{based on observations obtained at Rozhen
National Astronomical Observatory, Bulgaria and at Hoher List, Germany}}

\author{
V. Stanishev\inst{1}$^{\star\star}$
\and
Z. Kraicheva\inst{1}$^{\star\star}$
\and
H.M.J. Boffin\inst{2}$^{\star\star}$
\and
V. Genkov\inst{1}\thanks{
e-mail: {\tt zk@astro.bas.bg}
(ZK), {\tt henri.boffin@oma.be} (HMJB), {\tt nao@mail.orbitel.bg} (VG)}}

\institute{Institute of Astronomy, Bulgarian Academy of Sciences,
           72, Tsarighradsko Shousse Blvd., 1784 Sofia, Bulgaria
   \and
    Royal Observatory of Belgium,
              Avenue Circulaire 3, B-1180 Brussels, Belgium
             }

\authorrunning{V. Stanishev et al.}
\offprints{V. Stanishev, \\ \email{vall@astro.bas.bg}}

\date{Received ;accepted}

\abstract{Results of a photometric study of the SW Sex novalike
\object{PX And} are presented. The periodogram analysis of the
observations obtained in October 2000 reveals the presence of
three signals with periods of 0\fd142, 4\fd8 and 0\fd207. The
first two periods are recognized as "negative superhumps" and the
corresponding retrograde precession period of the accretion disk.
The origin of the third periodic signal remains unknown. The
observations in September-October 2001 point only to the presence
of ``negative superhumps" and possibly to the precession period.
The origin of the ``negative superhumps" is discussed and two
possible mechanisms are suggested. All light curves
 show strong flickering activity and power spectra with a typical red
noise shape. PX And shows eclipses with highly variable shape and
depth. The analysis suggests that the eclipse depth is modulated
with the precession period and two possible explanations of this
phenomenon are discussed.
An improved orbital ephemeris is also determined:
$T_{\rm min}[HJD]=49238.83662(14)+0\fd146352739(11)E$.
}
\maketitle

\section{Introduction}

\object{PX And} is probably one of the most complicated SW Sex stars
(Thorstensen et al. \cite{th91}; Hellier \& Robinson \cite{hel94};
Still et al. \cite{still}).
Thorstensen et al. (\cite{th91}) reported shallow eclipses with highly
variable eclipse depth and repeating with a period of $\sim$0\fd1463533.
The authors  assumed a steady-state accretion disk effective
temperature distribution $T_{\rm eff}\sim r^{-3/4}$ and $q\simeq0.46$,
 and adjusted the inclination and disk radius to match
the width and depth of the mean eclipse. They obtained
$i\simeq73.8^\circ$ and $r_{\rm d}\simeq0.6R_{L_1}$. No other
system parameters estimations have been published and all authors
used the above values. Apart from the distinctive characteristics
of SW Sex stars (reviewed recently by Hellier \cite{helrev}),
\object{PX And} shows some other interesting peculiarities.
Patterson (\cite{patt99}) reported \object{PX And} to show
simultaneously ``negative" and ``positive" superhumps, and signals
with typical periods of 4--5 days. This implies that \object{PX
And} most probably possesses an eccentric and tilted accretion
disk. In the view of the expected stream overflow (Hellier \&
Robinson \cite{hel94}) all this would result in very complex
accretion structures.

In this paper we report the results of a photometric study of \object{PX And}.

\section{Observations and data reduction}

 \begin{figure*}[t]
 \centering
 \includegraphics*[width=18cm]{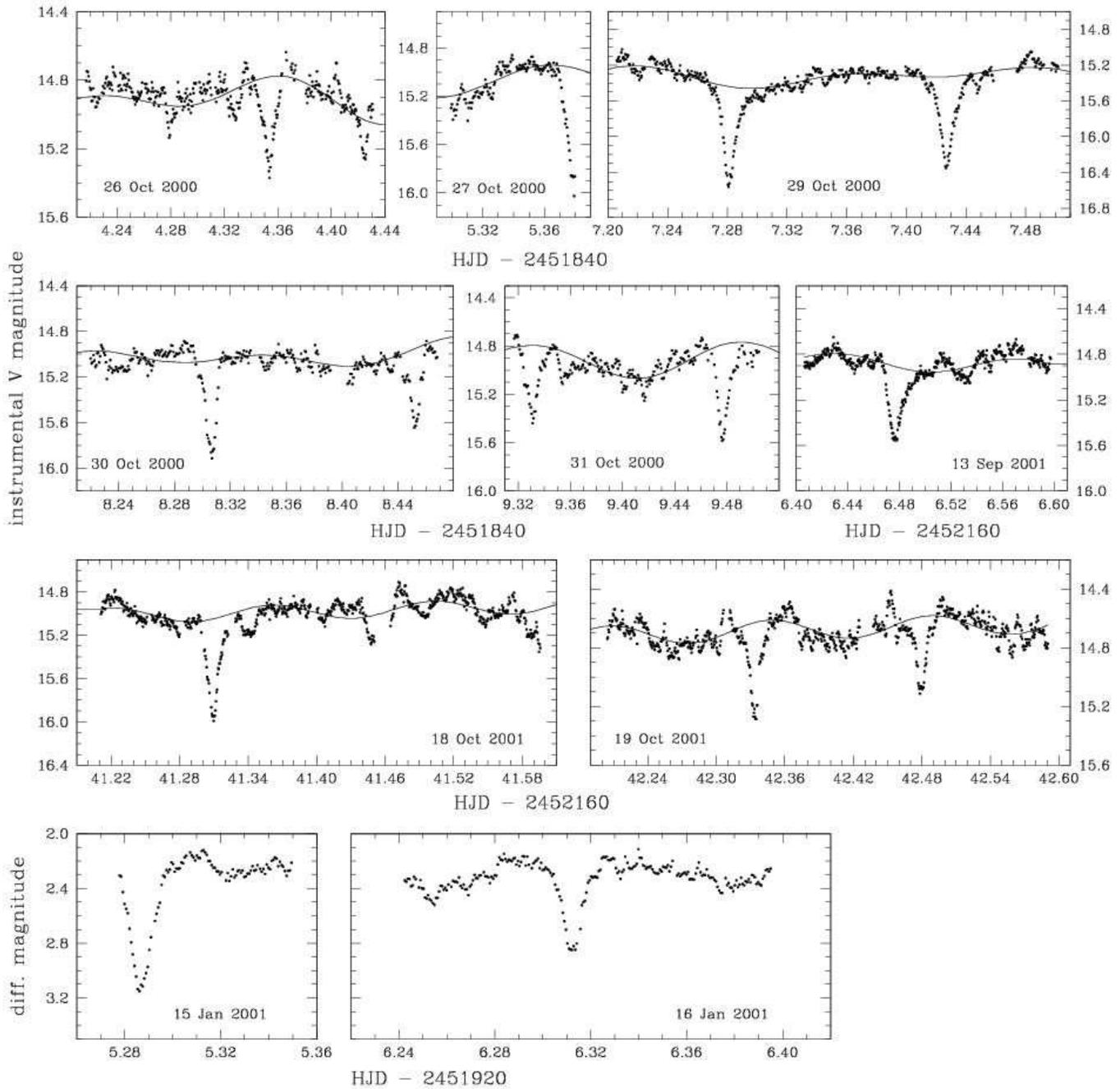}
 \caption{Observations of \object{PX And}. The best fits with the
 detected periods are overplotted. Note the different magnitude scale.}
 \label{pats}
 \end{figure*}

Photometric CCD observations of \object{PX And} were obtained with
the 2.0-m telescope of Rozhen Observatory. A Photometrics 1024$^2$
CCD camera  and a Johnson $V$ filter were used. The atmospheric
conditions were stable with seeing $\sim$2$^{''}$ and a 2$\times$2
binned camera was used. This resulted in $\sim3$ pixels per FWHM
and $\sim$13 s read-out dead-time. In total 8 runs were obtained
in 2000 and 2001. The exposure time used was between 20 and 40 s.
In addition, two unfiltered runs were obtained with the 1-m
telescope at Hoher List Observatory. Some details of the
observations are given in Table\,\ref{obs}. After bias and
flat-field corrections the photometry was done with the standard
DAOPHOT aperture photometry procedures (Stetson \cite{ste}). The
magnitude of \object{PX And} was measured relatively to the star
And1-5 ($V$=15.908) and And1-8 ($V$=16.613) served as a check
(Henden \& Honeycutt \cite{comp}). The runs are shown in
Fig.\,\ref{pats}.

\section{Results}

\subsection{The ephemeris}

The observations cover 14 usable eclipses. The corresponding
eclipse timings were determined by fitting Gaussian to the lower
half of the eclipses and are given in Table\,\ref{obs}. Combining
them with the eclipse timings published by Li et al. (\cite{li}),
Hellier \& Robinson (\cite{hel94}), Andronov et al. (\cite{andr})
and  Shakhvskoy et al. (\cite{shak}) we determined the following
orbital ephemeris:
\begin{equation}
T_{\rm min}[HJD]=49238.83662(14)+0\fd146352739(11)E.
\end{equation}

\subsection{Periodogram analysis}

\begin{table}[t]
\centering
\caption[]{CCD observations of PX And.}
\vspace{0.2cm}
\begin{tabular}{lccc}
\hline
\hline
\noalign{\smallskip}
 Date      & Start & duration  &  mid-eclipse  \\
                  &  HJD-2450000 & [hours] & [HJD]    \\

\hline
\noalign{\smallskip}
\multicolumn{4}{c}{Rozhen 2-m, Johnson $V$ filter }\\
\hline
\noalign{\smallskip}
2000 Oct 26  & 1844.22 & 5.13 & 1844.35372 \\
2000 Oct 27  & 1845.30 & 1.93 & 1845.37879: \\
2000 Oct 29  & 1847.21 & 7.10 & 1847.28135 \\
             &         &      & 1847.42712 \\
2000 Oct 30  & 1848.22 & 5.95 & 1848.30635 \\
             &         &      & 1848.45240 \\
2000 Oct 31  & 1849.33 & 4.50 & 1849.33129 \\
             &         &      & 1849.47647 \\
2001 Sep 13  & 2166.41 & 4.59 & 2166.47673 \\
2001 Oct 18  & 2201.21 & 9.23 & 2201.30935 \\
2001 Oct 19  & 2202.20 & 9.26 & 2202.33354 \\
             &         &      & 2202.47930 \\
\hline
\noalign{\smallskip}
\multicolumn{4}{c}{Hoher List 1-m, unfiltered }\\
\hline
\noalign{\smallskip}
2001 Jan 15  & 1925.28 & 1.72 & 1925.28667 \\
2001 Jan 16  & 1926.24 & 3.67 & 1926.31250 \\
\hline
\end{tabular}

\label{obs}
\end{table}

\subsubsection{Low-frequency periodicities}

\begin{figure}[t]
 \centering
 \includegraphics*[width=8.8cm]{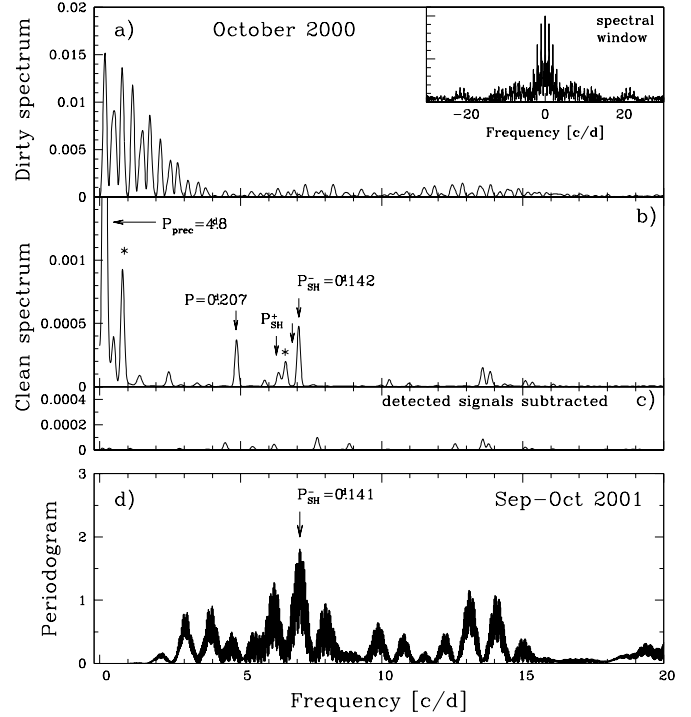}
 \caption{ {\bf a--c)} Periodogram analysis of the October 2000 runs. The
 detected periods are marked with the labelled arrows. The orbital
 period (the arrow without label) and the ``positive superhumps"
 period reported by Patterson(\cite{patt99}) are also indicated.
 The peaks marked with ``*" are artifacts from the cleaning and do
 not correspond to real periodicities in the data. {\bf d)} Lomb-Scargle
 periodogram of the October 2001 runs.}
 \label{per}
 \end{figure}

The mean out-of-eclipse magnitude of \object{PX And} shows large
variations. The five October 2000 runs obtained within 6 nights
suggest that the mean magnitude is modulated with a period of
$\sim$5 days and a full amplitude of $\sim$0.5 mag. After the
eclipses have been masked, the October 2000 data were searched for
periodic brightness modulations by computing the Fourier
transform. The complex spectral window of the data results in a
power spectrum which is completely dominated by the aliases of the
strong 5-days wave (Fig.\,\ref{per}$a$). The computed power
spectrum was deconvolved with the spectral window by using the
CLEAN algorithm (Roberts et al. \cite{rob}). The CLEANed spectrum
is shown in Fig.\,\ref{per}$b$. Apart from the strong
low-frequency peak there are a number of peaks remaining after the
cleaning. \object{PX And} has already been reported to show
``negative superhumps" with a period of 0\fd1415 and cycles with
periods of $4-5$ days (Patterson \cite{patt99}; Center for
Backyard Astrophysics (CBA) web site\footnote{ {\tt
http://cba.phys.columbia.edu/results/pxand}}). The origin of these
signals is not fully understood, but they are believed to be
caused by a retrograde precession of an accretion disk which is
tilted with respect to the orbital plane. In this model the
relation between the orbital period $P_{\rm orb}$, the ``negative
superhumps" period $P^-_{\rm SH}$ and the precession period of the
disk $P_{\rm prec}$ is given by
\begin{equation}
\frac{1}{P_{\rm prec}}=\frac{1}{P^-_{\rm SH}}-\frac{1}{P_{\rm orb}}.
\label{prec}
\end{equation}
If the peak at $P\simeq0\fd142$ is associated with the ``negative
superhumps", then the expected precession period of the disk is
$\sim$4\fd8. Thus, the low-frequency wave seen in the data is most
likely a manifestation of the disk precession. We note that
 CLEAN is an iterative algorithm, which is initiated with the
 assumption that the strongest peak in the power spectrum corresponds
 to a real periodicity in
the data. Thus, to apply CLEAN algorithm to our \object{PX And}
data we have to assume that the $\sim$5-days modulation is real.
Although the baseline of our observation is short, they strongly
indicate
 the presence of a $\sim$5-days modulation.  In addition,
CBA observations also reveal such a modulation. Thus, our
assumption seems reasonable.

To search for additional weaker signals, the $\sim$0\fd142 and
$\sim$4\fd8 waves were subtracted from the data by performing a
non-linear least-squares  multi-sinusoidal fit with unknown
amplitudes and phases. With observations over a baseline of only 6
days, $P_{\rm prec}$ could not be accurately determined; because
of this $P_{\rm prec}$ was calculated from Eq.\,(\ref{prec}).  The
CLEANed spectrum of the residuals (not shown) shows that all
strong peaks are removed.  The only exception is the peak at
$P\simeq0\fd207$. The appearance of this peak is quite unusual
because it could not be related to any clock in a system with
$P_{\rm orb}$=0\fd146352739. This peak should be regarded with
caution since the corresponding period is roughly equal to the
mean duration of the runs. Peaks corresponding to periods of the
order of the duration of the runs or harmonics of a period of 1
day may appear in the power spectrum because of the color
difference between the target and the comparison star. The full
amplitude of the 0\fd207 signal is however $\sim$0.15 mag; too
large to be caused by the second-order extinction. Moreover, the
second-order extinction in $V$ band is negligible. To check if
this peak is not an artifact of the cleaning we applied several
ways of pre-whitening. We subtracted fits with the $\sim$4\fd8 and
$\sim$0\fd142 periods only, a fit with both periods but allowing
them to vary in a narrow interval around peak center, etc. The
results show that the peak at $P\simeq$0\fd207 is always present.
We further estimated what is the probability for this peak to
appear in the power spectrum by chance alone. This estimate is
complicated by the fact that the noise in \object{PX And} light
curves is red rather than white (Sec.\,\ref{sec:high}). Because of
this we applied the following procedure. We have generated 1000
artificial red noise time series sampled exactly as the
observations, applied CLEAN to each of them and then estimated the
probability to see peaks as strong as the peak at $\sim$0\fd207 by
chance. The artificial time series were generated by the method
suggested by Timmer \& K\"onig (\cite{tim}). This method allows
one to generate random realizations of a given red noise process.
The only input information needed is the power spectrum of the
process. We used the mean red noise power spectrum shape derived
in Sec.\,\ref{sec:high}, which
 ensures that the artificial time series have on average the same
power spectrum as the \object{PX And} light curves. The analysis
of the CLEANed spectra of the artificial time series shows that
the peak at 0\fd207 is statistically significant by more than
99\%. Further, we examined whether the peak at 0\fd207 could be
introduced in the CLEANed spectrum by the presence of the other
two periodic signals, in conjunction with the red noise and the
gaps in the data. To do this we added to the artificial red noise
time series two sinusoids with periods of 4\fd8 and 0\fd142, and
amplitudes and phases as determined from the fit of the original
data. None of the CLEANed spectra of these time series show strong
peaks around 0\fd207. This suggests that the presence of the 4\fd8
and 0\fd142 signals could not be the reason for the appearance of
the 0\fd207 peak in the CLEANed spectrum of \object{PX And} data.

 \begin{figure}[t]
 \centering
 \includegraphics*[width=8.8cm]{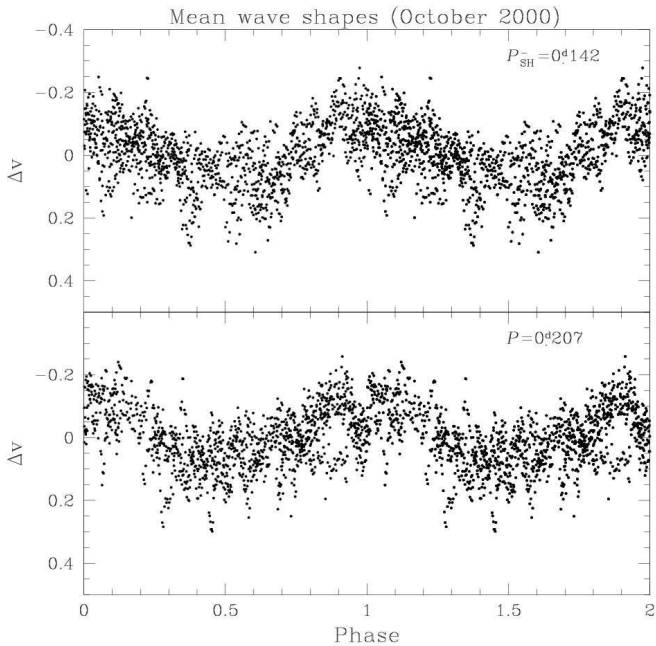}
 \caption{The October 2000 light curves folded with the 0\fd142 and
 0\fd207 periods. Before the folding, the remaining two periodic
 signals were subtracted from the data.}
 \label{mean}
 \end{figure}

In order to refine the periods we again performed a non-linear
least-squares  multi-sinusoidal fit to the data, but this time the
periods were also allowed to vary in a narrow interval around the
peaks maxima. $P^-_{\rm SH}$ and $P_{\rm prec}$ are not
independent, so in the function used to fit the data they were
coupled through Eq.\,(\ref{prec}). Thus, only $P^-_{\rm SH}$ and
$P_1$=0\fd207 were varied. In that way the following periods were
determined: $P^-_{\rm SH}$=0\fd142\,$\pm$0.002,
$P_1$=0\fd207\,$\pm$0.004 and $P_{\rm prec}$=4\fd8 with
semi-amplitudes of  0.086, 0.076 and 0.256 mag, respectively. The
errors of $P^-_{\rm SH}$ and $P_1$ are estimated from the
half-width at half-maximum of the central peak of the spectral
window. A reliable error estimate of the 4\fd8 period is not
possible mainly because the data set is short and covers only
$\sim$1 cycle. The best fit is shown in Fig.\,\ref{pats} and the
mean wave shapes of the "negative superhumps" and 0\fd207 signal
in Fig.\,\ref{mean}. The CLEANed spectrum of the residuals
(Fig.\,\ref{per}$c$) shows no peaks which implies that all peaks
in the power spectrum are successfully accounted for by the fit
with the detected periods.

Periodogram analysis was performed to the three runs obtained in
2001, after the subtraction of the mean of each run. The
Lomb-Scargle periodogram (Scargle \cite{scar}) is shown in
Fig.\,\ref{per}$d$. The sparse distribution of the runs resulted
in a complex alias structure of the periodogram. It is clear,
however, that the strongest peak corresponds to a ``negative
superhump" with $P^-_{\rm SH}\simeq$0\fd141. This value of
$P^-_{\rm SH}$ gives $P_{\rm prec}\simeq$3\fd8. The fit with these
two periods is shown in Fig.\,\ref{pats}. The corresponding
semi-amplitudes are 0.069 and 0.21 mag. The 0\fd207 period is not
detected in 2001. Note, however, that in both runs obtained on 18
and 19 October there is a slight increase of the out-of-eclipse
magnitude, which might be a manifestation of a periodicity around
0\fd207.

 \begin{figure}[t]
 \centering
 \includegraphics*[width=8.8cm]{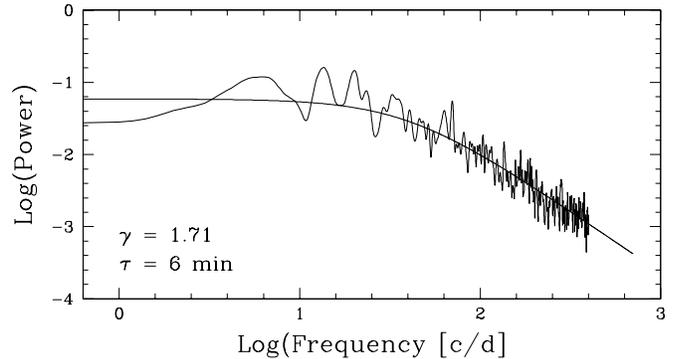}
 \caption{Double logarithmic scale plot of the mean power
 spectrum after the subtraction of the
 low-frequency signals. Also shown is
 the best fit with Eq.\,(\ref{flick}).}
 \label{psll}
 \end{figure}

\subsubsection{High-frequency periodicities}
\label{sec:high}

 \begin{figure*}[!t]
 \sidecaption
 \includegraphics*[width=12cm]{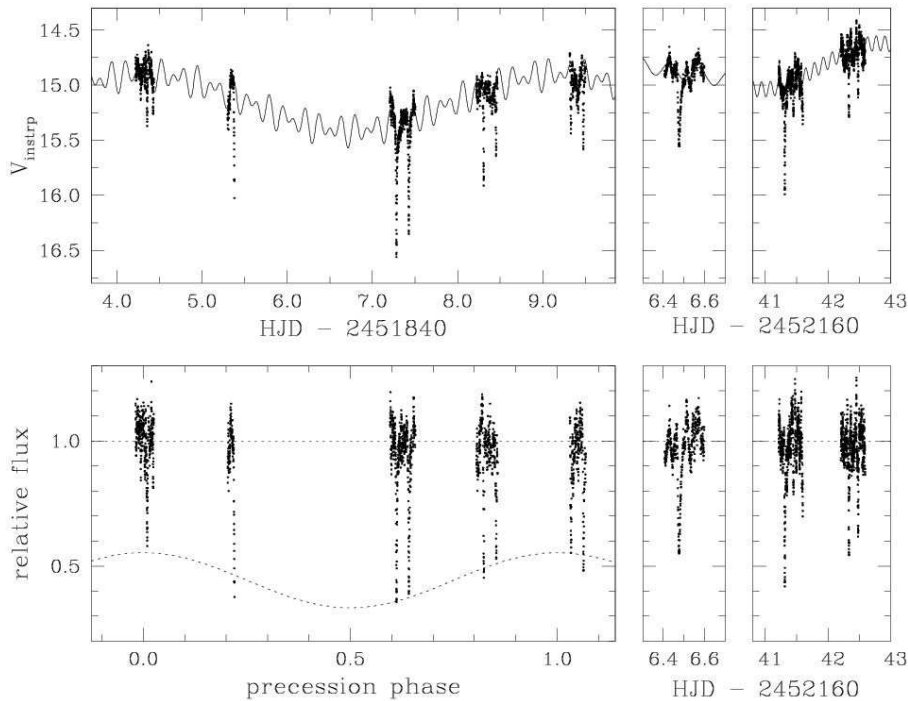}
 \caption{{\it Upper row:} $V$ band observations of PX And and the
 best fits with the detected periods. {\it Lower row:} The same
 data, but with the best fit subtracted and displayed in flux
 scale. The fit to the eclipse depths is also shown.}
 \label{eclch}
 \end{figure*}

To search for high-frequency periodicities we subtracted the
low-frequency signals from the data and calculated the power
spectra. The periodograms show many peaks, but no coherent
oscillations were detected. In double logarithmic scale the power
decreases linearly with frequency (Fig.\,\ref{psll}) and this is
usually interpreted as a result of the flickering (Bruch
\cite{bruch}). The power spectrum can be described by the equation
\begin{equation}
P(f)=\frac{\alpha}{1+(2\pi\tau f)^\gamma}
\label{flick}
\end{equation}
with $\gamma$ being the slope of the linear part and  $\tau$ may
serve as an estimation of the mean duration of the dominant
structures in the light curves. Equation\,(\ref{flick}) was fitted
to the mean power spectrum yielding $\gamma$=1.71 and $\tau$=6
min. We also determined $\tau$ by means of autocorrelation
function as described in Kraicheva et al. (\cite{krmv}). The mean
value of $\tau\simeq$4~$\pm$1.5 min is slightly lower, but
consistent with that determined from the fit of the mean power
spectrum. The mean standard deviation in the light curves after
the subtraction of the low-frequency signals is 0.061~$\pm$0.008
mag. This value is consistent with the standard deviation found in
the light curves of \object{TT Ari} during the ``negative
superhumps" regime (Kraicheva et al. \cite{krtt2}). After
switching to ``positive superhumps", the standard deviation in the
\object{TT Ari} light curves decreased by a factor of 2. It would
be interesting to compare this quantity in the novalikes showing
only ``negative" or ``positive superhumps" to see if there is a
systematic difference.

 \begin{figure}[t]
 \centering
 \includegraphics*[width=8.8cm]{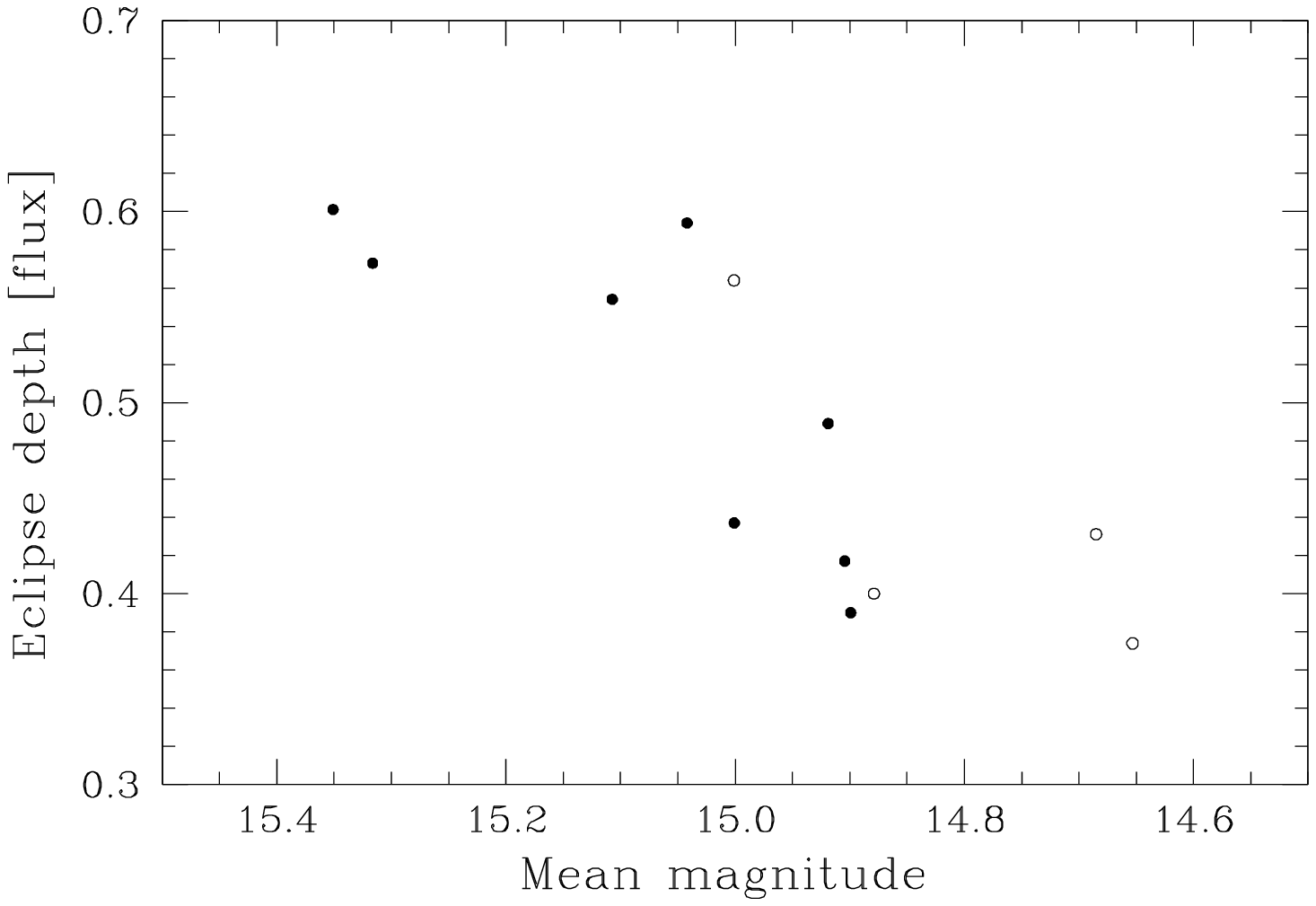}
 \caption{Eclipse depth as a function of the out-of-eclipse
 magnitude. Dots and open circles show the 2000 and 2001 data,
 respectively.}
 \label{pxdepth1}
 \end{figure}

\subsection{Variable eclipse depth}

As first noticed by Thorstensen et al. (\cite{th91}) the eclipse
depth in \object{PX And} is highly variable. The October 2000 runs
leave the impression that the eclipse depth is modulated with the
precession cycle (Fig.\,\ref{eclch} upper row). This however could
be a visual effect: the eclipses look deeper when they coincide
with the superhumps minima. To verify this we have subtracted the
best fit from the data and show the residuals in relative flux in
the lower row of Fig.\,\ref{eclch}. The eclipse depth modulation
with the precession period is evident even after the subtraction.
Figure\,\ref{pxdepth1} shows the eclipse depth as a function of
the mean out-of-eclipse magnitude. We fitted the eclipse depths
with a sinusoid with the precession period. This shows that the
eclipses are deepest at the minimum of the precession cycle. The
mean eclipse depth is $\sim$0.56 ($\sim$0.89 mag) and the
amplitude of the variation is $\sim$0.11. Thus, the eclipses we
see are on average much deeper than the value of 0.5 mag reported
by Thorstensen et al. (\cite{th91}).  In Figs.\,\ref{eclch} and
\ref{pxdepth1} one can also find hints
 that in 2001 \object{PX And} was $\sim$0.2 mag brighter than in 2000.

\section{Discussion}

The origin of the ``negative superhumps" is still an open
question. The most plausible model is based on the assumption of a
retrograde precession of a tilted accretion disk. While the
"positive superhumps" have been simulated numerically since the
work of Whitehurst (\cite{wh}), there are some difficulties to
simulate the "negative superhumps". In fact, all attempts to
simulate ``negative superhumps" failed in the sense that they were
not able to produce a significant tilt starting from a disk lying
in the orbital plane (Murray \& Armitage \cite{mur}; Wood et al.
\cite{wood}). Once tilted, however, the accretion disk starts
precessing in retrograde direction (Larwood et al. \cite{lar};
Wood et al. \cite{wood}) thus giving two additional clocks:
$P_{\rm prec}$ in inertial co-ordinate system and $P^-_{\rm SH}$
in a system co-rotating with the binary. $P_{\rm prec}$ is
typically a few days and thus $P^-_{\rm SH}$ is slightly shorter
than the orbital period. The angle at which the accretion disk is
seen from the Earth is modulated with the precession period,
giving brightness modulations with $P_{\rm prec}$.  Taking into
account foreshortening and limb-darkening, one can estimate the
disk tilt needed to produce the observed amplitude of the 4\fd8
modulation in \object{PX And}. With a limb-darkening coefficient
$u=0.6$ the tilt angle is between 2.5$^\circ$ and 3$^\circ$,
depending on the assumed system inclination. For comparison the
simulations of Wood et al. (\cite{wood}) were performed with a
tilt angle of 5$^\circ$.

 \begin{figure}[t]
 \centering
 \includegraphics*[width=8.8cm]{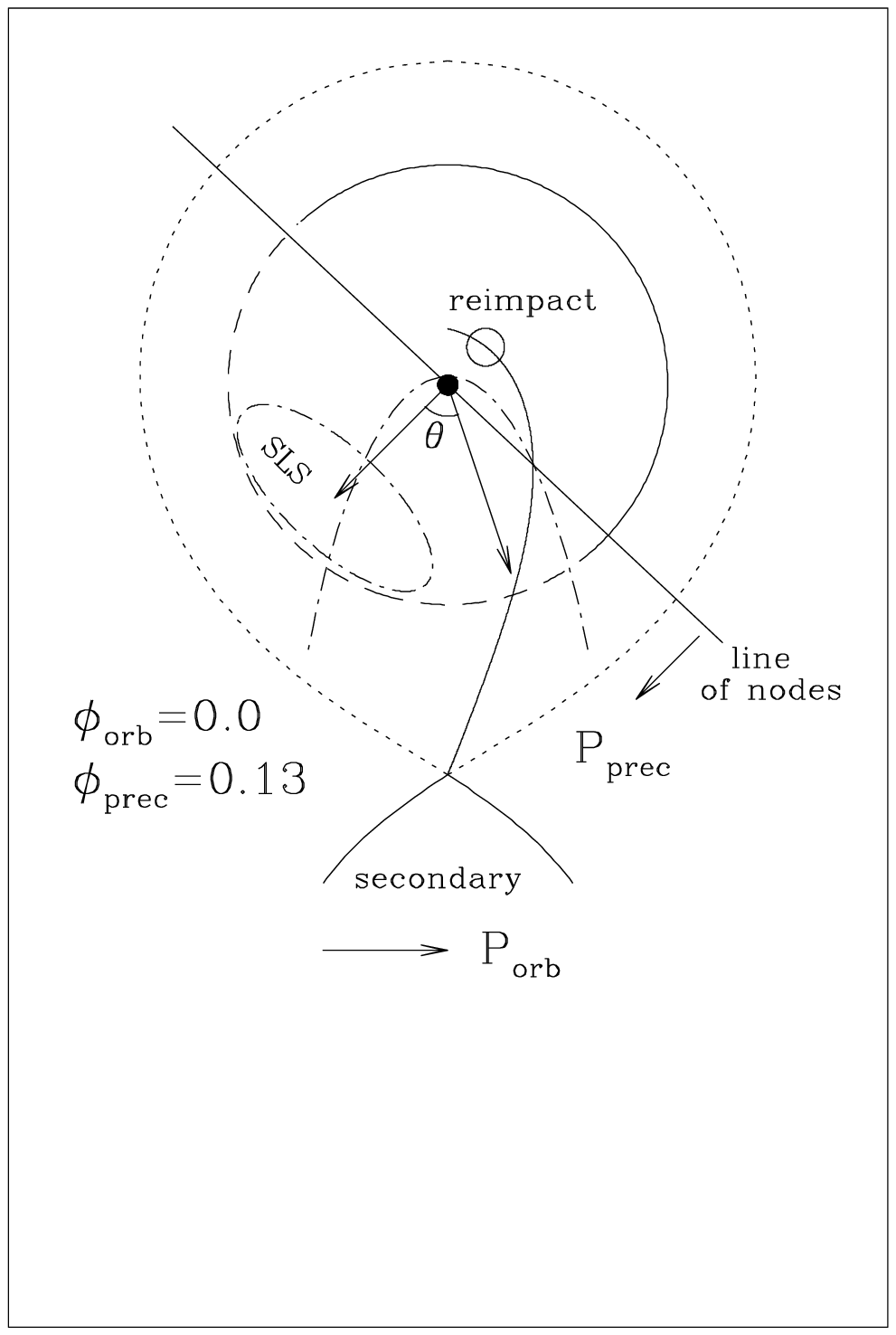}
 \caption{A sketch of the system configuration at $\phi_{\rm
 prec}$=0.13 {\it and} $\phi_{\rm orb}$=0. The disk edge marked with the
 dashed line lays below the orbital plane. Also shown are the
 expected position of the SLS, shadow of the secondary, accretion
 stream and re-impact zone.}
 \label{pxmod}
 \end{figure}

The mechanism generating the ``negative superhump" light itself is
however rather uncertain. According to the simulations of Wood et
al. (\cite{wood}) the two opposite parts of the disk which are
most displaced from the orbital plane are tidally heated by the
secondary. The heating is maximal when these parts point to the
secondary and this happens twice per superhump cycle. Wood et al.
(\cite{wood}) find that the tidal stress is asymmetric with
respect to the disk mid-plane and the disk surface facing the
orbital plane is more heated. Thus, if the disk is optically thick
the observer sees only one side and a modulation with the
superhump period is observed. Let us define precession phase
$\phi_{\rm prec}$ to be zero at the maximum of the 4\fd8 cycle,
i.e. when the disk is most nearly face-on. The relative phasing
between the periodic signals determined from the fit shows that
the maxima of the superhumps coincide with the eclipses at
$\phi_{\rm prec}\simeq$0.13. Figure\,\ref{pxmod} shows a sketch of
the system configuration in eclipse at $\phi_{\rm prec}=0.13$.
 The expected position of the superhump light source
(SLS) according to Wood et al. (\cite{wood}) and the line of nodes
are also shown. The dashed line marks the part of the disk which
lays below the orbital plane. According to the simulations of Wood
et al. (\cite{wood}) the "negative superhumps" maxima  occur when
the region labelled as "SLS" in Fig.\,\ref{pxmod} lies on the line
connecting the two system components, which means that the
superhumps maxima will coincide with the eclipses at $\phi_{\rm
prec}=0$. Our results show that in \object{PX And} this is not
exactly the case and the superhumps maxima are observed
$\sim0.13P_{\rm SH}^-$ later.

There is another mechanism which could generate ``negative
superhumps" and at the same time account for the delay. Patterson
(\cite{patt99}) noticed that most of the SW Sex novalikes show
``negative superhumps" and this could naturally explain the
accretion stream overflow thought to be responsible for the SW Sex
phenomenon (Hellier \& Robinson \cite{hel94}). In a system with a
precessing tilted accretion disk, the amount of gas in the
overflowing stream will vary with the ``negative superhumps"
period. Correspondingly, the intensity of the spot (shown in
Fig.\,\ref{pxmod}) formed where the overflowing stream re-impacts
the disk should be modulated and might be the ``negative" SLS. A
similar model was proposed by Hessman et al. (\cite{hess}) for the
origin of "positive superhumps". The maximal overflowing is
expected to take place when the most displaced part of the disk
points to the accretion stream. The intensity of the re-impact
spot will however reach its maximum later when the gas in the
stream reaches the re-impact zone (the moment shown in
Fig.\,\ref{pxmod}).  This means that the gas in the accretion
stream which at the moment shown in Fig.\,\ref{pxmod} is near the
re-impact zone, has passed over the disk edge earlier, when the
system configuration was different and the most displaced part of
the disk pointed to the accretion stream.

Let the gas in the overflowing stream travels from the outer disk
edge to the re-impact zone in the time $\Delta t$. Then, because
in co-ordinate system rotating with the binary the line of nodes
precesses with period $P^-_{\rm SH}$,
 at the superhumps maxima the displaced part
of the disk will not point to the accretion stream but will be
rotated with respect to it by an angle $\theta\simeq360\Delta
t/P^-_{\rm SH}$ deg. Figure\,\ref{pxmod} shows that in the case of
\object{PX And} $\theta\simeq0.15\times360=54^\circ$. Therefore,
in order to observe the maxima of the ``negative superhumps" at
$\phi_{\rm prec}\simeq$0.13 {\it and} $\phi_{\rm orb}\simeq$0 the
travel time from the outer disk edge to the re-impact zone has to
be $\sim30$ min. This is slightly longer than expected (Warner \&
Peters \cite{war}), but we have to take into account that the
velocity of the incoming gas is most probably reduced by the first
impact with the accretion disk. This model for the origin of the
``negative" SLS can be independently checked by tracking the
intensity of the high-velocity $s$-wave observed in the spectra of
SW Sex novalikes.

The origin of the 0\fd207 signal is quite uncertain. The
corresponding peak in the power spectrum could not be a result of
an amplitude modulation of the ``negative superhumps". If this is
the case, the power spectrum should show two additional peaks. The
secondary crosses the line of nodes twice per superhump cycle. The
time between three consecutive passages coincides within 1.5\%
with the observed period. It is however not clear how this
would produce periodic brightness modulations, moreover with the
observed amplitude of $\sim$0.15 mag.

The most puzzling observation of \object{PX And} is its highly
variable eclipse depth. Our observations suggest that the eclipse
depth is modulated with the precession period (Figs.\,\ref{eclch}
and \ref{pxdepth1}), indicating that this phenomenon could be
related to the disk precession. In the precessing tilted accretion
disk model, in some particular combinations of $q$ and $i$ the
eclipsed area of the disk is expected to vary with the precession
period and thus to modulate the eclipse depth. We have simulated
eclipses of a tilted precessing disk which showed that the eclipse
depth indeed varied but with a very low amplitude of 2-3\%, which
is insufficient to explain the observations.

 \begin{figure}[t]
 \includegraphics*[width=8.8cm]{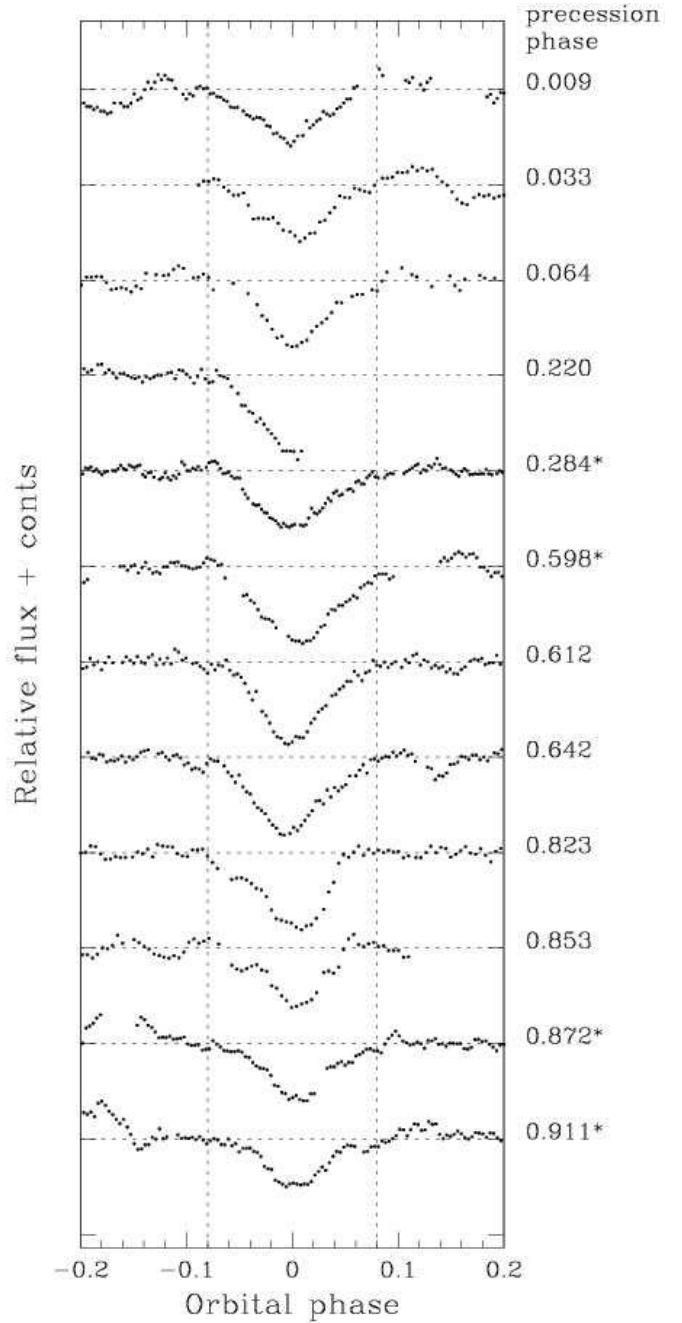}
 \caption{Normalized $V$ band eclipses of PX And. ``*" mark the
 eclipses obtained in 2001.}
 \label{ecl}
 \end{figure}

The relatively shallow eclipses  show that the accretion disk in
\object{PX And} is not totally eclipsed. In this case, if there is
a significant constant light source which is totally eclipsed, the
eclipse depth will be modulated since different fraction of the
total light emitted by the system will be eclipsed at different
precession phases. The deepest eclipses are expected at the
precession cycle minima, as observed. As the constant light source
must be eclipsed, its most likely location is close to the white
dwarf. The white dwarf itself could not be the source of the
constant light as it does not contribute significantly to the
total system light. The most obvious candidate is the inner hot
part of the disk. Since the brightness modulation with the
precession cycle comes from more or less pure geometrical
considerations it is not clear why the emission of the inner parts
of the disc could be constant. One possible explanation might be
that the inner part of the disk is not tilted or is less tilted
than the outer part. In this  case the emission from the inner
disk will be constant. Another source of the constant light could
be the emission from the boundary layer between the disk and the
white dwarf surface. This however is not very likely since the
temperature of the boundary layer is thought to be rather high to
emit significantly in the optical wavelengths. Of course, there
exists the possibility that the eclipse depth modulation with the
precession cycle is an artifact. If some flickering peaks occur
during the eclipse (which is likely because the accretion disk in
\object{PX And} is not totally eclipsed), this will decrease the
observed eclipse depth. Figure\,\ref{ecl} shows that some eclipses
are indeed badly affected by the flickering. Because of the small
number of eclipses covered this might introduce a spurious
modulation. But since the correlation between eclipse depth and
mean magnitude is good, we suggest that the eclipse depth
modulation with the precession cycle is real.

 \begin{figure}[t]
 \includegraphics*[width=8cm]{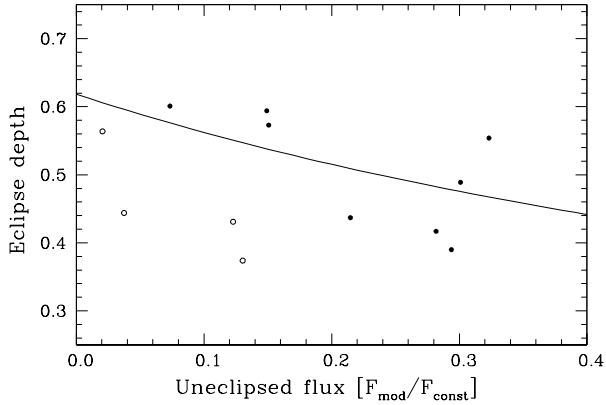}
 \caption{Eclipse depth as a function of $F_{\rm mod}/F_{\rm
 const}$ (see text for details). Dots and open circles show the
 2000 and 2001 data, respectively.}
 \label{pxdepth}
 \end{figure}

There is another way to modulate the eclipse depth. As the SLS is
non-uniformly distributed over the disk surface and the accretion
disk is not totally eclipsed, there are two extreme possibilities
when the SLS is (i) totally and (ii) never eclipsed. In both cases
the eclipse depth will depend on the actual superhump light at the
moment of eclipse, but will have opposite behavior. If the SLS is
totally eclipsed, then the deepest eclipses will be observed when
superhump maxima coincide with the eclipses. If the SLS is never
eclipsed, the eclipse depth will be affected in a way analogous to
the {\it veiling} of the absorption lines in \object{T Tau} stars.
 In these stars the additional emission in the continuum reduces
the observed depth of the absorption lines. Then the eclipses will
be deepest when superhump minima coincide with the eclipses. Using
the parameters of the periodic signals obtained from the
multi-sinusoidal fit, we calculated the total contribution of the
modulated part of the flux to the unmodulated one, $F_{\rm
mod}/F_{\rm const}$, at the moments of the eclipses. $F_{\rm
const}$ is modulated with the precession cycle and hence may be
thought to be constant on the superhump time scale. The
$\sim$0\fd207 signal was included in the calculations for 2000.
The reason is that the eclipse depth variation observed in the
2000 data is rather high to be explained if only the ``negative"
SLS is involved. Figure\,\ref{pxdepth} shows the eclipse depth
$d_{\rm e}$ as a function of $F_{\rm mod}/F_{\rm const}$. Although
the scatter is large, one should keep in mind that any flickering
near the mid-eclipses will decrease the eclipse depth and will
move the points in Fig.\,\ref{pxdepth} downward. If a small number
of eclipses are observed then this could easily obscure any
correlation. Despite the large scatter, there are some hints of an
inverse correlation in Fig.\,\ref{pxdepth}. This suggests that the
SLS is not eclipsed. In Fig.\,\ref{pxdepth} is also shown an
arbitrary curve defined by the equation:
\begin{equation}
d_{\rm e}=\frac{d_{\rm e,0}}{1+F_{\rm mod}/F_{\rm const}}
\label{depth}
\end{equation}
which gives the eclipse depth as a function of the additional
modulated light in the case  when this light is not eclipsed.
Here, $d_{\rm e,0}$ is the eclipse depth if $F_{\rm mod}=0$.

Although mathematically the veiling could explain the eclipse
depth changes, there are many uncertainties with this
interpretation. Most of them are related to the unknown place of
origin of the periodic signals. If the conclusion of Wood et al.
(\cite{wood}) about where the ``negative superhumps" are generated
is correct, then the ``negative" SLS should be at least partially
eclipsed. This suggests that the stream overflow could be the
mechanism generating the ``negative superhumps" in \object{PX
And}. Moreover, in a system with grazing eclipses the re-impact
zone is not eclipsed. Apart of this, the origin of the
$\sim$0\fd207 signal is unknown and it is not clear if its source
is eclipsed or not. Thus, the question why the eclipse depth in
\object{PX And} varies remains open and we have only pointed to
some of the possible explanations.

Thorstensen et al. (\cite{th91}) have estimated $q$, $i$ and the
disk radius of \object{PX And} by fitting a mean eclipse. It
should be noted however that the observations of Thorstensen et
al. (\cite{th91}) point to an average eclipse depth of $\sim0.5$
mag while our observations show much deeper eclipses. It is clear
that the highly variable shape and depth of eclipses could affect
significantly any attempt to estimate the system parameters of
\object{PX And}. Because of this  we have not attempted such an
estimation from the eclipse profiles. It would also not be correct
to define so called "mean" eclipse and to analyze it with the
eclipse mapping technique. Moreover, a large part of the accretion
disk in \object{PX And} is not eclipsed. The eclipse mapping
algorithm could be easily modified to allow analysis of tilted
accretion disks. To use this technique however one will need to
average at least several eclipses at given precession phase in
order to reduce the influence of flickering and other noise.

Generally, if a given cataclysmic variable shows ``positive
superhumps" one could estimate $q$ using the existing
$\epsilon^+(q)$ relations ($\epsilon^+=(P_{\rm sh}^+-P_{\rm
orb})/P_{\rm orb}$). Patterson (\cite{patt99}) reported that
\object{PX And} shows ``positive superhumps" with
$\epsilon^+\simeq$0.09 and this could be used to estimate $q$.
Recently, Montgomery (\cite{mon}) published an analytic
$\epsilon^+(q)$ expression which takes into account the pressure
effect. Applying this relation  (Eq.\,(8) in Montgomery
\cite{mon}) to \object{PX And} we obtain $q\simeq$0.27. Although
this value of $q$ is more plausible for a system showing "positive
superhumps", one should keep in mind that the peak in the power
spectrum of the CBA data is very weak. The "negative superhumps"
are much more confidently detected, but unfortunately an
$\epsilon^-(q)$ relation for the "negative superhumps" has not
been established yet. Patterson (\cite{patt99}) noticed that
$\epsilon^-\simeq-0.5\,\epsilon^+$. Our periodogram analysis
yields $\epsilon^-\simeq-0.029$, hence $\epsilon^+\simeq0.058$.
From Montgomery (\cite{mon}) relation, one obtains $q\simeq0.18$.
This is not an unrealistic value but we cannot be sure that the
relation $\epsilon^-\simeq-0.5\,\epsilon^+$ holds true for
\object{PX And}.

Based on the analysis of our new photometric observations
 and on the results of Patterson (\cite{patt99})
we can with no doubt place the novalike \object{PX And} among the
cataclysmic variables showing permanent "negative superhumps". The
observations of the star  in 2000 show  two unusual
 features: the presence of another superhump with a period of
$\sim$0\fd207 and a modulation of the eclipse depth with the
precession period of the accretion disk. Our 2001 observations are
however limited in time coverage and it is difficult to say if
these two phenomena are typical of the star or are a peculiarity
of this particular data set only. As a suggestion  for future work
we recommend performing a multi-site photometric observations
covering several precession cycles. If the modulation of the
eclipse depth with the precession period is confirmed, this
phenomenon could be further studied by analysis of the eclipse
profiles at different precession phases.

\begin{acknowledgements}
We are grateful to the referee John Thorstensen for his valuable
comments and suggestions. HB wishes to thank Prof. Wilhelm
Seggewiss for generously allocating time at Hoher List. The work
was partially supported by NFSR under project No.~715/97.
\end{acknowledgements}

\end{document}